\documentclass[aps,pre,twocolumn]{revtex4-1}

\usepackage{longtable}
\usepackage{graphicx}
\usepackage{dcolumn}
\usepackage{bm}
\usepackage{amsmath}
\usepackage[psamsfonts]{amssymb}
\usepackage{subfigure}
\usepackage{slashbox}

\begin{document}

\title{Six-fold crystalline anisotropic magnetoresistance in the (111) LaAlO$_3$/SrTiO$_3$ oxide interface}



\author{P. K. Rout, I. Agireen, E. Maniv, M. Goldstein, and Y. Dagan}
\email[]{yodagan@post.tau.ac.il}
\affiliation{Raymond and Beverly Sackler School of Physics and Astronomy, Tel-Aviv University, Tel Aviv, 69978, Israel}


\date{\today}

\begin{abstract}
We measured the magnetoresistance of the 2D electron liquid formed at the (111) LaAlO$_3$/SrTiO$_3$ interface. The hexagonal symmetry of the interface is manifested in a six-fold crystalline component appearing in the anisotropic magnetoresistance (AMR) and planar Hall data, which agree well with symmetry analysis we performed.
The six-fold component increases with carrier concentration, reaching 15\% of the total AMR signal. 
Our results suggest the coupling between higher itinerant electronic bands and the crystal as the origin of this effect and demonstrate that the (111) oxide interface is a unique hexagonal system with tunable magnetocrystalline effects.
\end{abstract}



\maketitle

\emph{Introduction.---}
The two dimensional electron liquid formed at the (100) interface between the two nonmagnetic insulators LaAlO$_3$ (LAO) and SrTiO$_3$ (STO) \cite{Hwang} features numerous properties, such as superconductivity \cite{Reyren, Caviglia1} and spin-orbit coupling \cite{BenShalom1,Caviglia2}, which are tunable by a gate voltage.
Past studies have shown evidence for magnetic order (possibly co-existing with superconductivity) whose exact character is not yet clear \cite{Brinkman,Seri,Sachs,Dikin,Li,Bert,Kaliski0,Salman,Lee,Ron},
prompting much theoretical activity \cite{Michaeli,Banerjee,Joshua,Chen}.

In-plane anisotropic magnetoresistance (AMR) can be employed as a probe for the magnetic properties of 2D structures, since it is sensitive to spin texture and spin-orbit interaction \cite{Trushin}. In the absence of crystalline anisotropy, rotating the magnetic field in the 2D plane results in a standard two-fold symmetric AMR term that depends on the angle between the magnetic field and the current. Such dependence has been observed in (100) LAO/STO \cite{BenShalom, Flekser, Fete}. It is absent in nonpolar doped STO heterostructure \cite{Flekser} and can hence be related to the Rashba spin-orbit interaction which is expected to be less important in the latter. It has been suggested that an easy axis for magnetization can be observed in the AMR \cite{Joshua}, however, the almost square symmetry of the interfacial crystal structure makes it difficult to distinguish between the two-fold term and the crystalline one. Here we explored a different interface, which has a hexagonal in-plane symmetry, namely the (111) LAO/STO heterostructure.

The stacking of (111) perovskite ABO$_3$ layers is AO$_3/$B$/$AO$_3/$B [Fig.~1(a)], and therefore with alternating -3e and +3e charges in LaAlO$_3$, whereas -4e and +4e in SrTiO$_3$ \cite{Herranz1}. In addition to the different polar structure compared to the (100) interface, its hexagonal symmetry has been predicted to be a key ingredient for the realization of various nontrivial states \cite{Xiao, Doennig}. Recently, six-fold Fermi contour related to the symmetry of STO (111) surface has been observed by angle resolved photoemission \cite{Rodel,Walker}. However, a distinct signature of six-fold symmetry of the (111) LAO/STO interface is lacking. In this Rapid Communication we show theoretically and experimentally that a six-fold crystalline $\cos(6\theta)$ term can be observed in the AMR of the (111) LAO/STO interface. This term depends on carrier concentration, suggesting that coupling between the lattice and higher energy itinerant electronic bands is an important ingredient for explaining magnetic effects in oxide interfaces.

\begin{figure}
\begin{center}
\includegraphics[width=1\hsize]{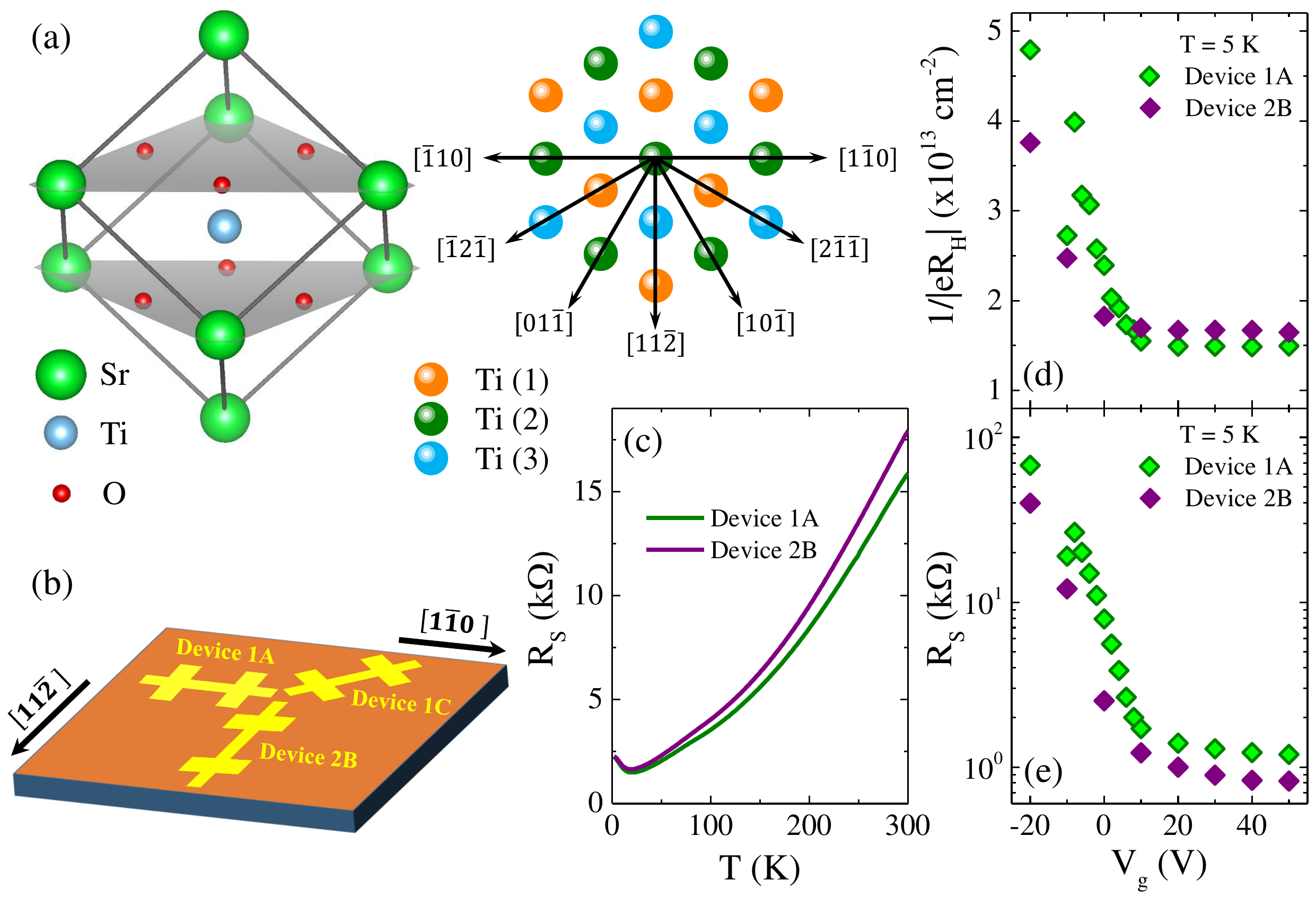}
\caption{(a) Left: Schematic depiction of SrTiO$_3$ unit cell showing (111) (SrO$_3$)$^{4-}$ planes. Right: Top view of three consecutive (111) Ti$^{4+}$ planes showing various in-plane crystal directions. The Ti atoms in three different planes are marked as Ti(1) (orange), Ti(2) (green), and Ti(3) (blue). (b) Schematic of different 100 $\mu$m $\times$ 260 $\mu$m Hall bar devices oriented along [1$\bar 1$0] (device 1A) and [11$\bar{2}$] (device 2B). Device 1C lies at an angle of 45$^{\circ}$ to the other two. (c) Temperature dependent sheet resistance $R_S$($T$) at $V_g = 0$~V. (d),(e) Gate dependence of the inverse Hall coefficient  1/$\left| {eR_H} \right|$ and sheet resistance $R_S$ at $T = 5$~K.\label{Figure1}}
\end{center}
\end{figure}


\emph{Methods.---}
Epitaxial thin films of LAO were deposited on atomically flat Ti-terminated STO (111) substrate (substrate preparation and structural characterisation are described in the Supplemental Material (SM)~\cite{Supple}) using pulsed laser deposition in oxygen partial pressure of $1\times$ 10$^{-4}$~Torr at 780$^{\circ}$C. After deposition the samples were annealed at 400$^{\circ}$C in 0.2 Torr oxygen pressure for one hour. We followed a three step deposition process~\cite{Maniv} to fabricate Hall bars oriented along different directions [Fig. 1(b)]. The thickness of LAO was 18 LaO$_3$/Al layers (each layer is $\approx$0.219~nm thick).

\par
\emph{Results.---} The typical sheet-resistance $R_{S}(T)$ as a function of temperature for two different devices (1A and 2B) is shown in Fig.~1(c). Both devices exhibit similar metallic behavior with a residual resistance ratio $R_S$(300~K)/$R_S$(10~K) of 10 and resistance values of $\sim$1.8~k$\Omega$ at 10~K. Below 18~K both devices show a small upturn in $R_{S}(T)$, similar to Refs.~\cite{Biscaras, Brinkman}. The sheet resistance $R_{S}$(5~K) decreases sharply when the gate voltage ($V_g$) is raised from -20~V to 10~V  [Fig.~1(e)]. Upon increasing $V_g$ further (up to 50~V) a much slower decrease in $R_{S}$ is observed decreases below 10~V. We have also extracted the gate dependent Hall coefficients from low field Hall measurements [Fig.~1(d)]. The sign of the Hall voltage is consistent with electron-like carriers. Surprisingly, 1/$\left| {eR_H} \right|$ decreases with an increase in $V_g$ up to 10~V and eventually saturates for higher $V_g$. This is in contrast to the expected behavior of increasing electron density with increasing  $V_g$. However, the Hall coefficients can not be directly used to determine carrier concentration ($n_S$) in oxide interfaces where the conduction is controlled by multiple $d$ bands interacting with each other \cite{Maniv}. Another interesting feature for (111) interface is that we only see monotonic decrease in  1/$\left| {eR_H} \right|$ with $V_g$ unlike nonmonotonic behavior observed in (100) or (110) interfaces \cite{Maniv,Biscaras2,Herranz2}. These observations are consistent with a recent study on (111) interface in Ref.~\cite{Davis}.

\begin{figure}
\begin{center}
\includegraphics[width=1\hsize]{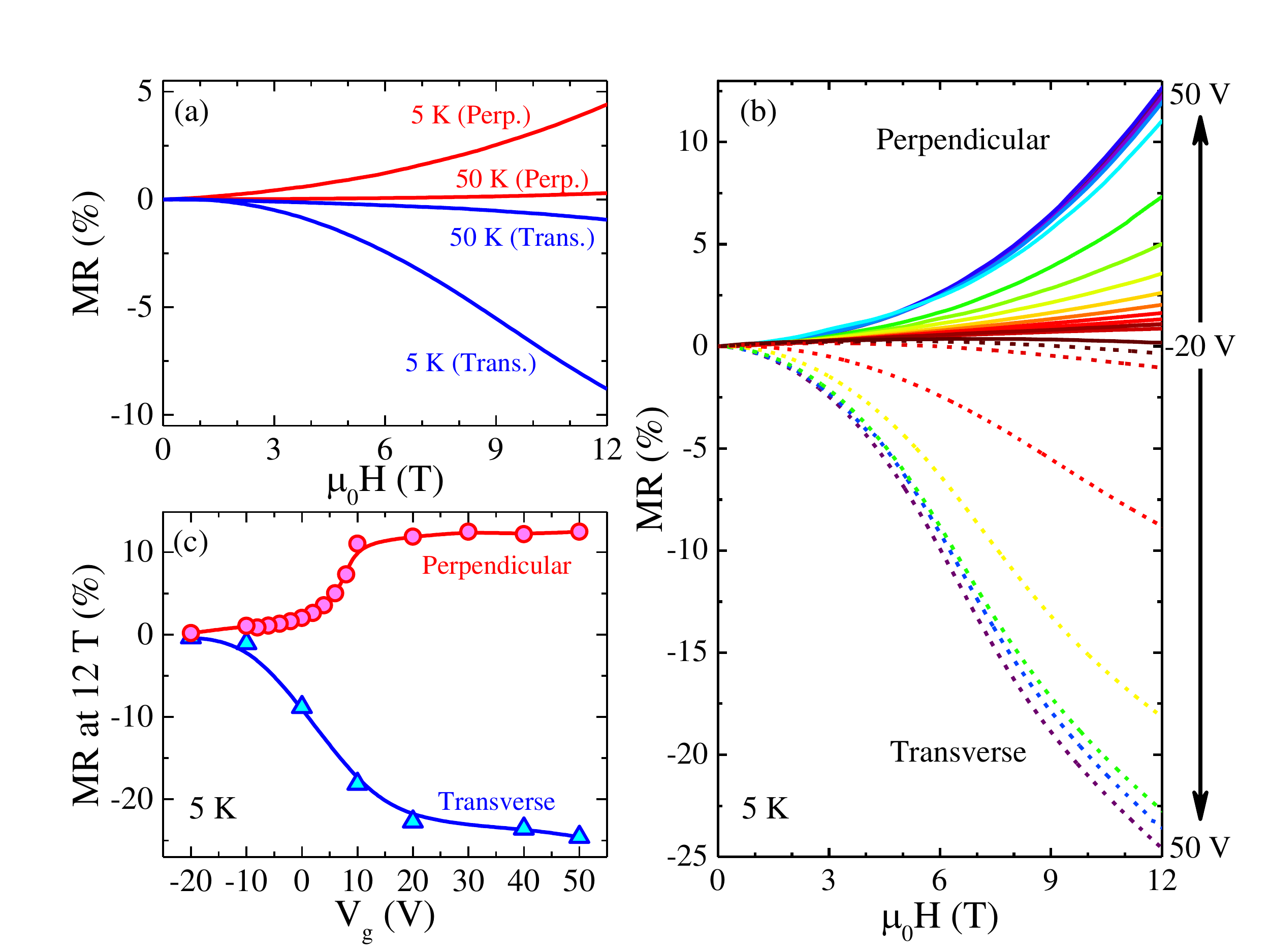}
\caption {Magnetoresistance for device 1A. (a) perpendicular ($\vec{H}$ perpendicular to $\vec{J}$ and interface) and transverse ($\vec{H}$ perpendicular to $\vec{J}$ and parallel to interface) magnetoresistance, MR = [R(H)-R(0)]/R(0), at 5~K and 50~K ($V_g=0$~V). (b)~MR as a function of $H$ at 5~K for different gate voltages. (c)~MR at 12~T and 5~K as a function of $V_g$.\label{Figure2}}
\end{center}
\end{figure}

Figure 2 shows the perpendicular ($\vec{H}$ perpendicular to $\vec{J}$ and interface) and transverse ($\vec{H}$ perpendicular to $\vec{J}$ and parallel to interface) magnetoresistance (MR) for device 1A. The data are symmetrized in the standard way. The orbital effect leads to positive perpendicular MR. By contrast, the transverse MR is negative, similar to Refs.~\cite{BenShalom,Diez}. The MR values decrease monotonically with decreasing $V_g$ as shown in Fig.~2(b). Both perpendicular and transverse MR values gradually rise with increasing gate until $V_g =$ 10~V and then saturate [Fig.~2(c)]. This saturation can be related to the saturation observed in $R_S (V_g)$ and 1/$\left| {eR_H}(V_g) \right|$ [Figs.~1(d) and 1(e)].

\par
We will now present the anisotropy in MR for two different angular rotations of the magnetic field. All these measurements are performed at both positive and negative magnetic fields (i.e., +13.5~T and -13.5~T). Figure 3(a) and 3(c) display the out-of-plane AMR$_{OP}$ for device 1A, which shows a maximum at $\psi =$ 90$^{\circ}$ and a sharper minimum at $\psi =$ 0$^{\circ}$. We observe a maximal AMR amplitude of $\approx$ 35 $\%$ for $V_g =$ 50~V, which starts decreasing with decreasing gate [Fig. 3(c)]. All these results are consistent with the MR data shown in Fig. 2. The in-plane AMR$_{IP}$ for device 1A reveals a maximum at $\theta =$ 0$^{\circ}$ ($\vec H \parallel \vec J$) and a minimum at $\theta =$ 90$^{\circ}$ ($\vec H \perp \vec J$) [see Fig. 3(b)]. At negative gate voltage values the observed AMR$_{IP}$ can be fitted with the standard twofold expression, $C\cos \left( {2\theta } \right)$. However, for positive gate voltages an increasing deviation from this two-fold behavior is observed [Fig.~3(d)].
\begin{figure}
\begin{center}
\includegraphics[width=1\hsize]{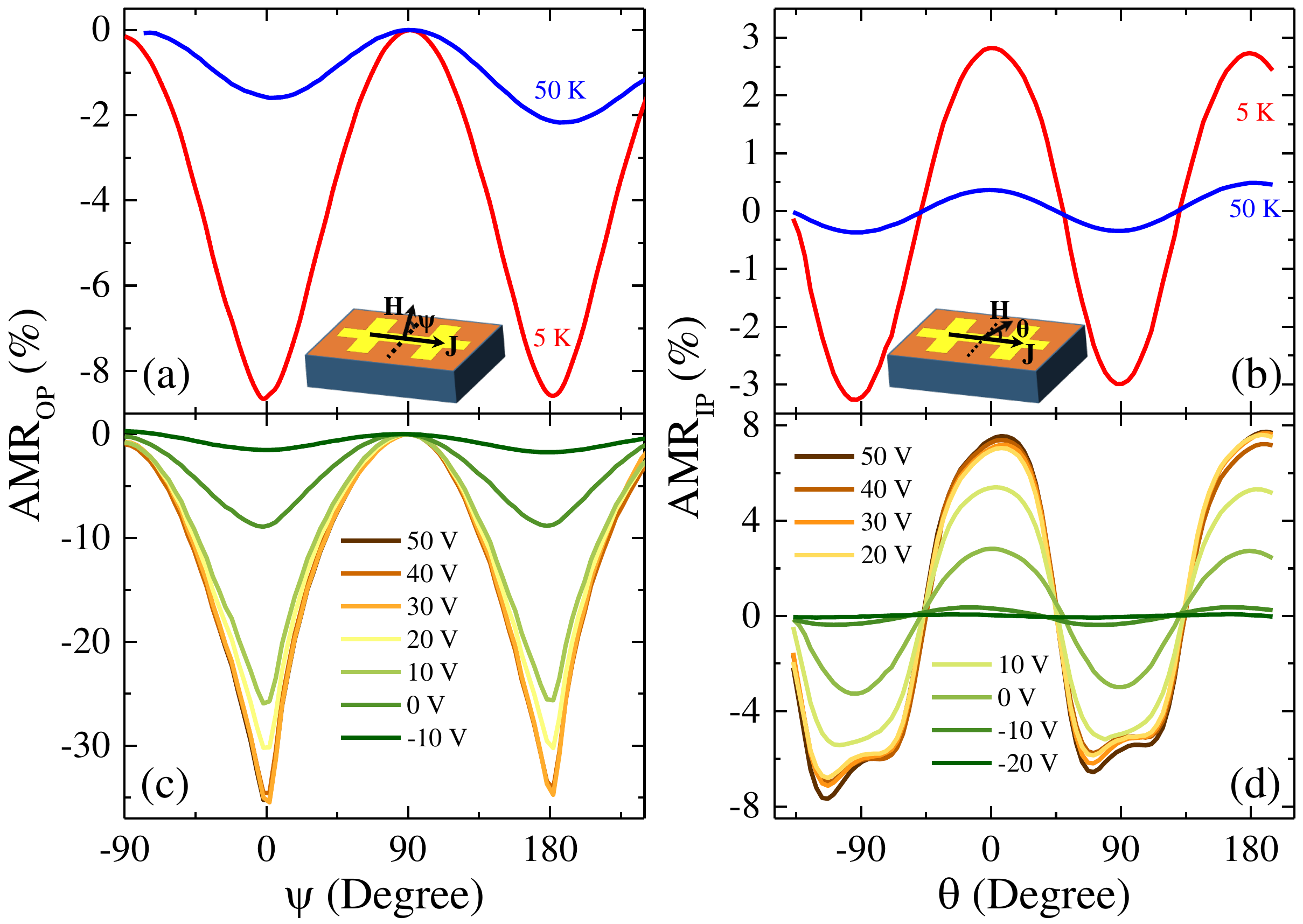}
\caption {AMR for device 1A. (a) Out-of-plane AMR$_{OP}(\psi)=[R(\psi)-R(0)]/R(0)$, at 5~K and 50~K for $V_g =$ 0~V. $\psi$ is the angle between $\vec H$ and the interface. (b) In-plane AMR$_{IP}(\theta)=[R(\theta)-R_{mean}]/R_{mean}$ at 5 K and 50 K for $V_g =$ 0 V. $R_{mean}$ is the angle-averaged resistance and $\theta$ is the angle between $\vec H$ and $\vec J$. (c),(d) Gate dependence of AMR$_{OP}(\psi)$ and AMR$_{IP}(\theta)$ at 5~K. For high temperatures (e.g., 50~K) and low gates (e.g., -20~V), AMR$_{IP}$ is maximal when $\vec{H} \parallel \vec{J}$. The corresponding angle is taken as $\theta =$ 0$^\circ$. \label{Figure3}}
\end{center}
\end{figure}
To understand this unusual behavior, we employ symmetry considerations to calculate the form of the AMR for a 2D hexagonal lattice system 
up to order $6\theta$ and find (see SM for details
\cite{Supple}):
\begin{equation}\label{Eq1}
\mathrm{AMR_{hex}} = {{C}_2} \cos (2\theta - 2\varphi ) +
{{C}_4} \cos (4\theta + 2\varphi )
 + {{C}_6}\cos (6\theta ),
\end{equation}
where $C_2$, $C_4$, and $C_6$ are two-, four-, and six-fold AMR coefficients, respectively. $\theta$ ($\varphi$) is the angle between $\vec H$ ($\vec J$) and the [1$\bar 1$0] crystal axis. To verify the role of the six-fold symmetry, we have tried to fit the AMR data for device 1A ($\varphi =$ 0$^\circ$) with only two- and four-fold components [$AMR_{Fit} = {C_2}\cos (2\theta ) + {C_4}\cos (4\theta )$], and found clear deviations for positive gate voltages [the case $V_g =$ 40~V is shown in Figs. 4(a) and 4(c)]. The residual data ($AMR_{IP}-AMR_{Fit}$) shows equally-spaced peaks separated by 60$^\circ$ [see Fig. 4(c)] --- a clear six-fold structure, which is well fitted by the last term of Eq.~(\ref{Eq1}). Interestingly, we observe an additional contribution to AMR at 30$^\circ$ (and presumably at -150$^\circ$, which however is outside our rotator's range). This feature appears even $V_g \le 0$, where the six-fold term is insignificant [Figs. 4(b) and 4(d)]. We discuss its origin below.
Figures 4(e) and (f) present the AMR coefficients as a function of gate voltage and temperature, respectively.
It should be noted that $|C_6|>|C_4|$, as expected for a hexagonal crystal structure. Moreover, $C_6$ appears at high gate voltages, and decays faster with temperature, as compared to $C_2$.

\begin{figure}
\begin{center}
\includegraphics[width=0.95\hsize]{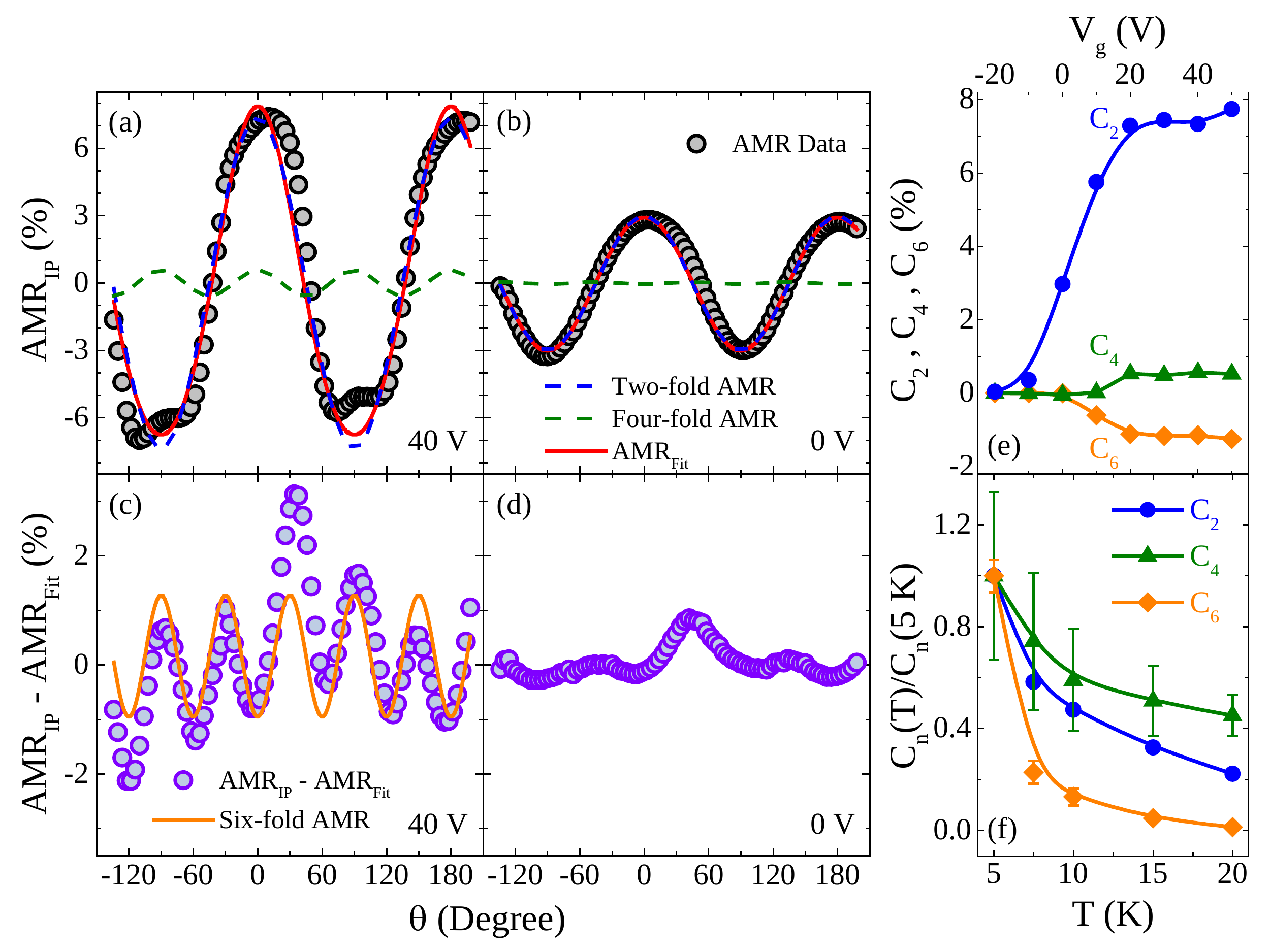}
\caption {AMR$_{IP}$ at 5 K for $V_g =$ 40~V (a) and 0~V (b)  measured on device 1A. The panels also include the two-fold and four-fold contributions, along with their sum, $AMR_{Fit}$. (c),(d) The same data after subtracting AMR$_{Fit}$ are shown along with six-fold fit. (e),(f) Gate voltage (at 5~K) and temperature (at 0~V, sample I4 \cite{Supple}) dependence of the AMR coefficients $C_2$, $C_4$, and $C_6$ [Eq.~(\ref{Eq1})]. \label{Figure4}}
\end{center}
\end{figure}

To further confirm that the AMR stems from the six-fold crystal structure, we measured it in two additional devices (1C and 1B), with different orientations with respect to the crystal axes ($\varphi =$ 45$^\circ$ and 90$^\circ$, respectively), and performed a similar fit (Fig.~5). The residual data for all three devices show similar six-fold behavior [see Figs.~5(d)-5(f)], which follows a ${C_6}\cos (6\theta )$ law, independent of $\varphi$, as expected for a 2D hexagonal system [Eq.~(\ref{Eq1})], indicating the crystalline nature of AMR. Our planar Hall data (Fig.~S2 in Ref.~\cite{Supple}) also corroborates this picture.
Let us note that such an angular dependence is inconsistent with square symmetry, as shown in the SM~\cite{Supple} (compare with Ref.~\cite{Ngai}).

Figure 5 also shows an additional uniaxial anisotropy lying along a crystal axis of the hexagonal system, similarly to Figs.~4(c) and 4(d). For devices 1A and 1C the anisotropy lies along the [2$\bar 1\bar 1$] direction, whereas it is along the [11$\bar 2$] direction for device 2B, as shown in Figs. 5(g)-5(i). The fact that these features can be seen at $V_g \le 0$~V, where the six-fold AMR is absent [Fig. 4(d)], implies that they have a different origin. One expects that the strong magnetic field of 13.5~T, at which our AMR measurements were carried out, will be able to align all magnetic domains (if these exist) along the magnetic field direction, thus ruling out magnetic domain structure as an explanation.
Another possibility is the presence of substrate terrace in Ti-terminated STO, which defines an anisotropy direction for conduction. However, the direction of the uniaxial component is not the same for our three devices, which were grown on the same substrate, thus eliminating this scenario as well.
All this implies that the observed uniaxial AMR anisotropy should be related to structural domains in STO. Recently, anisotropic conductance has been observed in (001) LAO/STO~\cite{Kalisky2,Frenkel}, which arises due to the formation of tetragonal domains in STO below 105~K. STO can have three different types of domains X, Y, and Z (distorted along the corresponding three principal directions of the original cubic lattice). These may form six boundary planes: (110) and (1$\bar 1$0) for X-Y, (011) and (0$\bar 1$1) for Y-Z, and (101) and (10$\bar 1$) for Z-X \cite{Honig}. These boundaries provide a low-resistance path, leading to anisotropy between the resistance in the parallel and perpendicular directions. In case of the (111) interface, these boundaries lie along [10$\bar 1$] and [$\bar 1$2$\bar 1$] for X-Y, [01$\bar 1$] and [2$\bar 1$$\bar 1$] for Y-Z, and  [1$\bar 1$0] and [11$\bar 2$] for Z-X. Therefore, we can have six possible anisotropy directions depending upon the type of domain boundaries. It is plausible that our devices situated at different regions of STO will have different domain boundaries, and hence different anisotropy directions. Furthermore, we note that sometimes after thermal cycling our devices above the cubic-to-tetragonal transition at 105 K and cooling back down the direction of the uniaxial anisotropy changes (see Fig. S7 in Ref.~\cite{Supple}). This gives further evidence that the uniaxial anisotropy in the AMR is a manifestation of the structural domains.

\begin{figure*}
\begin{center}
\includegraphics[width=0.7\hsize]{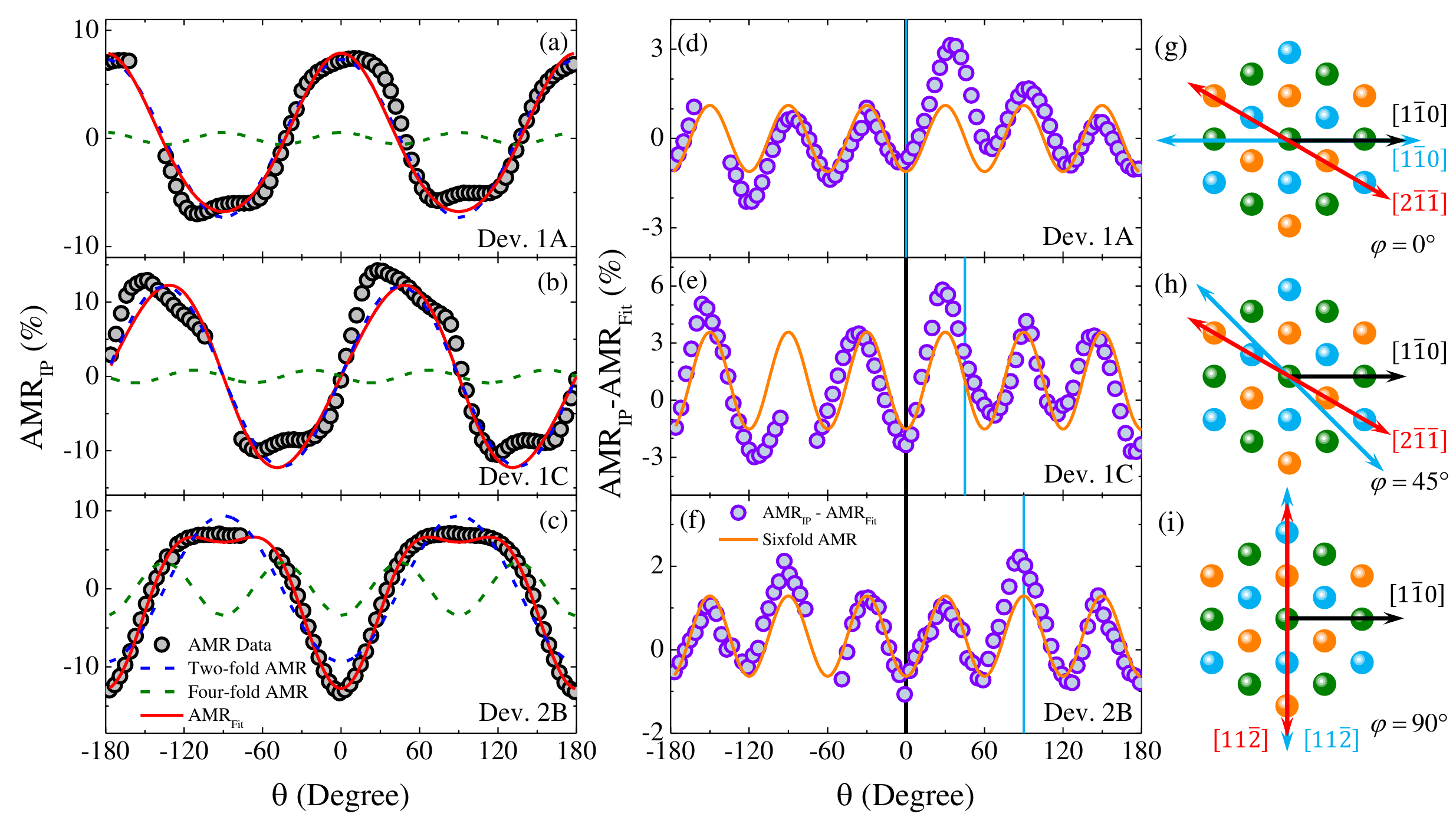}
\caption {(a)-(c) AMR$_{IP}$ for device 1A (a) measured at 5 K while for devices 1C (b) and 2B (c) at 2 K. The applied gate voltage is $V_g =$ 50 V. The solid line shows the $AMR_{Fit}$, which is composed of two-fold and four-fold AMR contributions (dashed lines). The values of fitting parameter $C_2$ are 7.7$\%$, 12.2$\%$, and 9.4$\%$ for devices 1A, 1C, and 2B, respectively, while the corresponding $C_4$ values are 0.56$\%$, 0.87$\%$, and 3.45$\%$. The data and fits are adjusted such that $\theta =$ 0$^{\circ}$ corresponds to the [1$\bar 1$0] direction. (d)-(f) The residual data ($AMR_{IP} - AMR_{Fit}$) for devices 1A (d), 1C (e), and 2B (f) along with six-fold AMR fits (solid line). The vertical black line represents the [1$\bar 1$0] direction while the current direction is shown by vertical blue line. (g)-(i) Schematic of (111) Ti$^{4+}$ planes showing the [1$\bar 1$0] crystal axis (black arrow), current direction (blue arrow), and extra uniaxial anisotropy direction (red arrow) for the three devices.\label{Figure5}}
\end{center}
\end{figure*}

Based on the above, one may then argue that by assuming that all six types of boundaries exist in equal proportions within the device, one could account for observed six-fold effect as well. However, as noted above, the six-fold effect and the uniaxial anisotropy appear in different gate voltage ranges. The sharp temperature dependence of the six-fold effect and its presence below 20 K [Fig.~4(f)] also disfavor structural domains as its origin. Finally, we have measured smaller (10~$\mu$m $\times$ 28~$\mu$m) Hall bar devices, and observed similar six-fold AMR (Fig.~S4 in Ref.~\cite{Supple}). Since existence of an approximately ``equal mixture'' of the six boundaries between two voltage terminals in such a small device is highly unlikely, we conclude that the six-fold AMR is a manifestation of the six-fold anisotropy of the (111) plane of the undistorted cubic crystal.

\emph{Discussion.---}
What could then be the origin of the six-fold crystalline AMR?
AMR necessitates spin-orbit interaction, and usually only appears in conjunction with magnetic order, typically ferromagnetic~\cite{McGuire} (see however Ref. \cite{Genish}).
In (100) LAO/STO, strong and tunable spin-orbit interaction has been reported \cite{BenShalom1,Caviglia2} and related to AMR effects \cite{Flekser,Fete}.
However, the evidence for magnetic order at the temperature range of our experiment ($T \ge$ 2~K) is elusive even for the better-studied (100) interface \cite{Brinkman,Seri,Sachs,Dikin,Li,Bert,Kaliski0,Salman,Lee,Ron}:
Magnetic hysteresis has been observed only below 1~K in a quasi-1D (100) system \cite{Ron}. Our current measurements on (111) LAO/STO have shown no indication of magnetic order (e.g., no anomalous Hall effect or hysteresis) in the temperature range studied.
Randomly-positioned magnetic dipoles have been observed in (100) LAO/STO \cite{Bert,Kaliski0}, which can account for the observations in the quasi-1D (100) case~\cite{Ron}. But it is unlikely that such scattered magnetic structure can account for the robust features we observe here with various device size and orientation.
\par
The AMR coefficients $C_2$ (standard AMR) and $C_6$ (crystalline AMR) have strikingly different temperature and gate voltage dependencies as well as opposite signs [Figs.~4(e) and 4(f)]. These findings suggest that they arise from different mechanisms whose microscopic details are not fully understood at this point. Now, we discuss various possibilities for the origin of the six-fold term.

The onset of six-fold crystalline AMR with $V_g$ surprisingly matches with the saturation observed in 1/$\left| {eR_H}(V_g) \right|$, suggesting a common mechanism for these two effects. In a (111) heterostructure the titanium t$_{2g}$ bands, which are believed to give rise to the 2D conduction, are split into two $J=$3/2 bands and a $J=$1/2 one \cite{Xiao,Doennig}. The hierarchy of these bands could not be predicted theoretically in these works, since it was found to sensitively depend on various intrinsic system parameters like strain and interactions. The dependence of $C_6$ on $V_g$ suggests that the higher electronic band couples strongly with the crystal sites. It would be natural to presume that the band with higher $J$ creates stronger AMR response.


\par
Recent theoretical studies on the (100) interface have shown a strong influence of spin-orbit interactions on crystalline AMR effects \cite{Fete,Diez,Bovenzi}. F\^{e}te \textit{et al.} have suggested that the AMR amplitude should be proportional to the square of $\Delta_{SO} / E_F$, where $\Delta_{SO} $ and $E_F$ are the Rasba spin-orbit coupling strength and the Fermi energy, respectively \cite{Fete,Raimondi}. Within this framework, one can conclude that $\Delta_{SO}$ increases rapidly as $V_g$ increases above $\sim$10 V and then saturates at higher $V_g$, reaching a value of order 10--15~meV, similar but somewhat higher values than those estimated for the (100) interface \cite{BenShalom1,Caviglia2,Fete}. Such a saturation is observed in many other transport properties of this interface.

\par
It should be mentioned that Joshua \textit{et al.} have reported that crystalline and noncrystalline AMR have different gate voltage dependence, and suggested that crystalline AMR is related to a Lifshitz transition in (100) LAO/STO \cite{Joshua} (see also Ref. \cite{Annadi}). On the other hand, by studying devices patterned along different crystal directions, Flekser \textit{et al.} have found that all the AMR features depend only on $\theta-\varphi$ (the angle between $\vec{J}$ and $\vec{H}$), indicating that AMR in the (100) system is noncrystalline \cite{Flekser}. This is very different from our current results for (111) LAO/STO, which have different dependence on $\theta$ and $\varphi$ [Fig.~5].

The AMR study on irradiated STO systems can also exhibit a six-fold AMR effect \cite{Ngai,Miao}. However, the sign of AMR is always opposite to the AMR observed in an LAO/STO interface (compare Ref. \cite{Ngai} to Ref. \cite{BenShalom} and Ref. \cite{Miao} to our results). This indicates that the AMR in irradiated systems results from orbital effects as suggested in Ref. \cite{Ngai}, while for LAO/STO interfaces, the important mechanism is spin scattering and spin-orbit interaction.


\emph{Conclusion.---}
To summarize, we utilize the unique hexagonal crystal symmetry of the (111) LAO/STO interface to provide conclusive evidence for the crystalline nature of AMR in such oxide interfaces. We establish that the magnetic effects in this interface are firmly related to underlying crystal structure. The strong dependence of AMR on carrier concentration suggests that the itinerant electrons from higher energy bands couple strongly to the lattice and contribute to the magnetic effects.

\emph{Acknowledgments.---}
This work has been supported by the Israel Science Foundation (Grant Nos. 569/13 and 227/15), the Israel Ministry of Science and Technology (contracts 3-11875 and 3-12419), and the Pazy Foundation (contract 268/17). P.K.R. acknowledges support from the Center for Nanosciences at Tel Aviv University. M.G. acknowledges the German-Israeli Science Foundation (Grant No. I-1259-303.10/2014), and the US-Israel Binational Science Foundation (Grant No. 2014262).

P. K. R. and I. A. contributed equally to this work.

\bibliographystyle{apsrev}

\clearpage

\newcommand{\beginsupplement}{%
        \setcounter{section}{0}
        \renewcommand{\thesection}{S\arabic{section}}%
        \setcounter{table}{0}
        \renewcommand{\thetable}{S\arabic{table}}%
        \setcounter{figure}{0}
        \renewcommand{\thefigure}{S\arabic{figure}}%
        \setcounter{equation}{0}
        \renewcommand{\theequation}{S\arabic{equation}}%
     }
     
\begin{widetext}

\section*{\textbf{Supplemental Material}}

In the Supplementary Material we provide details on our substrate preparation procedure (Sec.~S1) and the structural characterisation of the film (Sec.~S2). The expressions for the anisotropic magnetoresistance (AMR) and planar Hall effect (PHE) up to order $6\theta$ are derived for 2D systems with either hexagonal or square symmetry in Sec.~S3. We also present the PHE data (Sec.~S4), additional results on a large device in comparison with a small one (Sec.~S5), and the effect of thermal cycling (Sec.~S6).

\section{Substrate preparation}

\begin{figure}[b]
\begin{center}
\includegraphics[width=0.6\hsize]{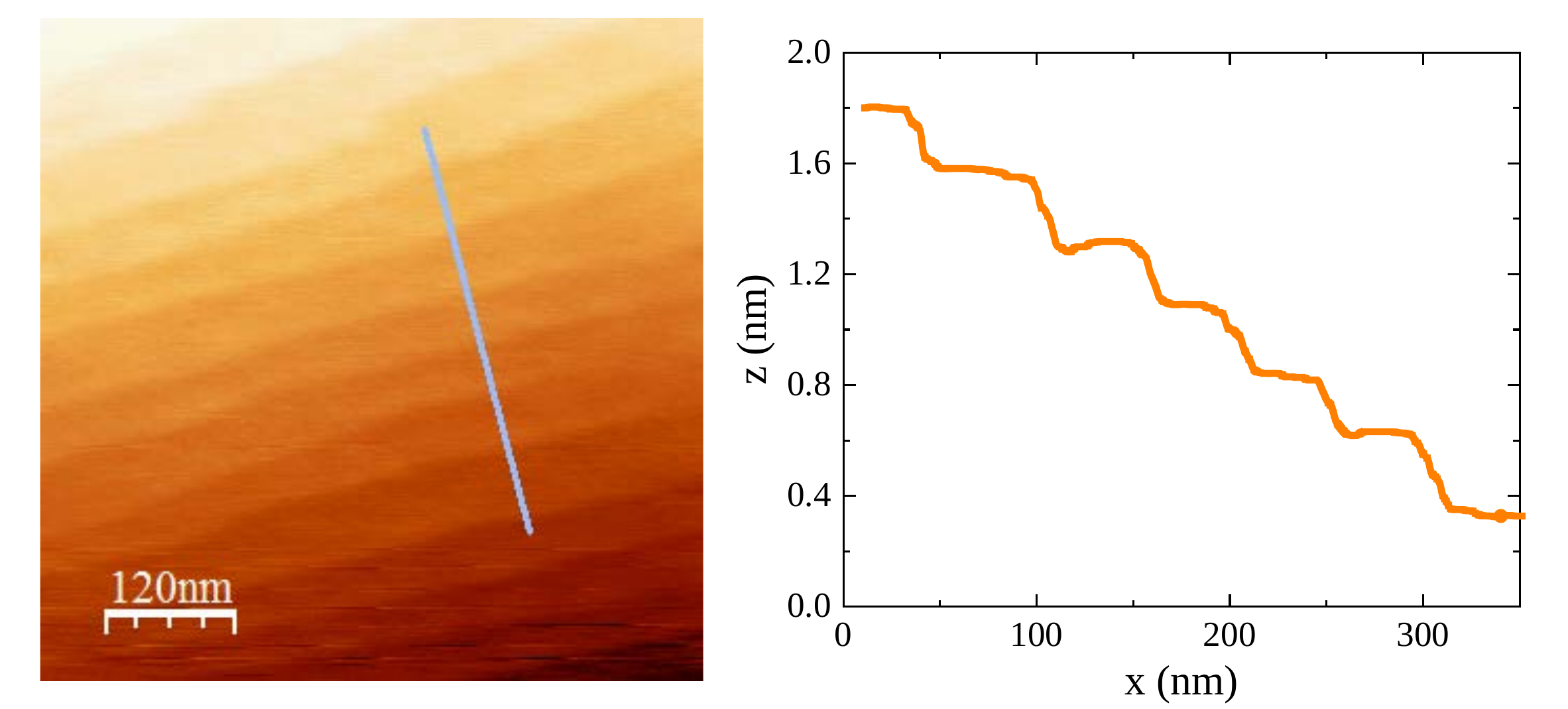}
\caption {Left panel: AFM image of an atomically smooth STO (111) surface displaying step and terrace structure. Right panel: A cut across the steps marked by the line in the AFM image. \label{FigureS1}}
\end{center}
\end{figure}

The surface treatment to obtain atomically flat Ti-terminated STO (111) consisted of combined chemical etching and annealing process. In the first step, unterminated 5 $\times$ 5~mm$^2$ STO (111) substrates were soaked in DI water for 30 minutes and subsequently annealed at 950$^\circ$C in oxygen for 3 hours. After soaking another 30 minutes in DI water with 10~min ultrasonic agitation, the substrates were dipped in buffered HF solution for 10~sec. Finally, the substrates were annealed at 950$^\circ$C in oxygen for 2~hours. Out of many trials, we were able to obtain few atomically smooth Ti-terminated substrates. Fig.~\ref{FigureS1} displays typical AFM image of a smooth substrate, which clearly shows the step and terrace structure with step heights of 2.2 ${\AA}$, matching the inter-planar distance of two consecutive Ti-layers in the [111]-direction.

\section{Structural Characterisation}
The film deposition was monitored by reflection high energy electron diffraction (RHEED). We observe RHEED streaks after film deposition [See the inset of Fig.~\ref{FigureS2}(a)], which indicates a 2D flat and crystallographically ordered surface. Fig.~\ref{FigureS2}(a) presents the $\theta$-2$\theta$ scan of LAO(8.8 nm)/STO (111) film showing the presence of (111) LAO peak close to (111) STO peak. A typical rocking curve ($\omega$-scan) about (111)  LAO peak [Fig.~\ref{FigureS2}(b)] has a full width half maximum of 0.3$^\circ$ indicating a high quality film growth.

\begin{figure}[t]
\begin{center}
\includegraphics[width=0.6\hsize]{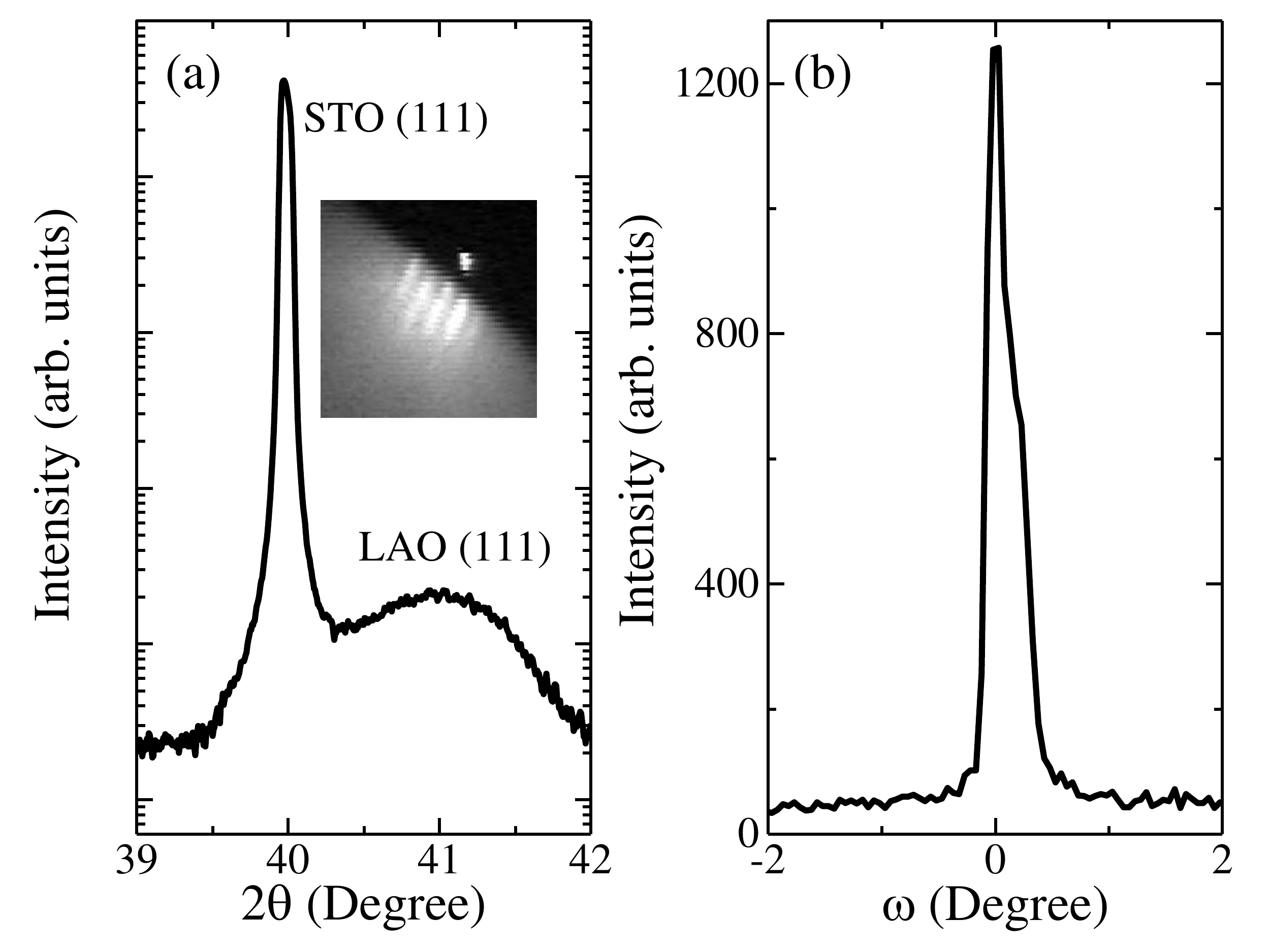}
\caption {(a) The $\theta$-2$\theta$ scan of LAO film around STO(111) peak. Inset: The RHEED images recorded along [1$\bar 1$0] after film deposition. (b) The rocking curve about (111) LAO peak. \label{FigureS2}}
\end{center}
\end{figure}

\section{Anisotropic magnetoresistance and planar Hall effect in square and hexagonal 2D crystals}

The symmetric part of the 2D resistivity tensor in the presence of an in-plane magnetic field directed along a unit vector $\vec{\alpha} = (\alpha_1, \alpha_2) \equiv (\cos\theta, \sin\theta)$ at an angle $\theta$ relative to some reference crystal axis, can be expressed as a series expansion in $\alpha_i$ \cite{Birss}:
\begin{equation}\label{Eq. S1}
{\rho _{ij}}(\vec {\alpha} ) = {a_{ij}} + {a_{klij}}{\alpha _k}{\alpha _l} + {a_{klmnij}}{\alpha _k}{\alpha _l}{\alpha _m}{\alpha _n} \\
 + {a_{klmnpqij}}{\alpha _k}{\alpha _l}{\alpha _m}{\alpha _n}{\alpha _p}{\alpha _q} + \dots.
\end{equation}
This tensor is even in the magnetic field, and hence, due to Onsager's relations, is symmetric in the indices $i,j$. By definition, each term in the expansion is symmetric under any interchange of the indices $k,l,\cdots$. Therefore, the value of each coefficient $a_{kl\cdots ij}$ only depends on how many of the indices $i,j$ and how many of the indices $k,l,\cdots$ equal 1 or 2 ($x$ or $y$, respectively).

For current $\vec J$ directed along the unit vector $\vec{\beta} = (\beta_1, \beta_2) \equiv (\cos\varphi, \sin\varphi)$ at an angle $\varphi$ relative to the reference crystal axis, the longitudinal resistance is given by
\begin{equation}\label{Eq. S2}
\rho_L (\overrightarrow \alpha  ,\overrightarrow \beta  ) = {\rho _{ij}}(\overrightarrow \alpha  ){\beta_i}{\beta_j},
\end{equation}
whereas the planar hall effect can be expressed as:
\begin{equation}
\rho_{PH} (\overrightarrow \alpha  ,\overrightarrow \beta  ) = {\rho _{ij}}(\overrightarrow \alpha  ){\beta^\prime_i}{\beta_j},
\end{equation}
where $\vec{\beta}^\prime = (-\beta_2, \beta_1)$ is a unit vector perpendicular to $\vec{J}$. We now present our results for these quantities in 2D square and hexagonal crystals up to order $6\theta$ (previous calculations only went up to $4\theta$ \cite{Birss}).


\subsection{2D square crystal}

For a 2D square crystal (space group \textbf{p4m} / point group $D_4$), the non-vanishing coefficients in Eq.~(\ref{Eq. S1}) and the relation between them are determined from symmetry analysis of $\rho _{ij}$ and are given in Tables~\ref{table C1}-\ref{table C4}.
Reflection about the $x$ ($y$) axis flips the 1 (2) 
component of electric field (polar vector; indices $i,j$), and the 2 (1) 
component of the magnetic field (axial vector; indices $k,l,\cdots$). Thus, by the \emph{mirror} symmetries, only coefficients with an even number of the indices $i,j,k,l,\cdots$ equal to 1 and even number equal to 2 do not vanish.
In addition, the 90$^\circ$ \textit{rotation} symmetry implies that interchanging the values of all the indices between 1 and 2 do not change the value of the coefficient; for example, $a_{11111212} = a_{12222212}$. Combining these findings with the fact that the order of the indices $i,j$ or $k,l,\cdots$ in Eq.~(\ref{Eq. S1}) is irrelevant, as discussed above, we find the results summarized in the Tables below.
\begin{table}[h]
\caption{\label{table C1} $a_{ij}$ from Eq.~(\ref{Eq. S1}) for 2D square and hexagonal crystals.}
\begin{ruledtabular}
\begin{tabular}{c|c c c c c c c c c}
$\beta_i\beta_j$ &  & $\beta_1^2$ &  & $\beta_2^2$ &  & $\beta_1\beta_2$ &  & $\beta_2\beta_1$ &  \\\hline
$ij$ &  & 11 &  & 11 &  &  &  & &  \\
\end{tabular}
\end{ruledtabular}
\end{table}
\begin{table}[h]
\vspace{-0.5cm}
\caption{\label{table C2} $a_{ijkl}$ from Eq.~(\ref{Eq. S1}) for 2D square and hexagonal crystals.}
\begin{ruledtabular}
\begin{tabular}{c|c c c c c c c c c}
\backslashbox{$\alpha_k\alpha_l$}{$\beta_i\beta_j$} &  & $\beta_1^2$ &  & $\beta_2^2$ &  & $\beta_1\beta_2$ &  & $\beta_2\beta_1$&  \\ \hline
$\alpha_1^2$ &  & 1111 &  & 1122 &  &  &  & &  \\
$\alpha_2^2$ &  & 1122 &  & 1111 &  &  &  & &  \\
$\alpha_1\alpha_2$ &  & &  & &  & 1212 &  & 1212 & \\
\end{tabular}
\end{ruledtabular}
\end{table}
\begin{table}[h!]
\vspace{-0.5cm}
\caption{\label{table C3} $a_{ijklmn}$ from Eq.~(\ref{Eq. S1}) for a 2D square crystal.}
\begin{ruledtabular}
\begin{tabular}{c|c c c c c c c c c}
\backslashbox{$\alpha_k\alpha_l\alpha_m\alpha_n$}{$\beta_i\beta_j$} & & $\beta_1^2$ & & $\beta_2^2$ & & $\beta_1\beta_2$ & & $\beta_2\beta_1$ & \\ \hline
$\alpha_1^4$ & & 111111 & & 111122 & &  & & & \\
$\alpha_2^4$ & & 111122 & & 111111 & &  & & & \\
$\alpha_1^3\alpha_2$ & & & & & & 111212 & & 111212 & \\
$\alpha_1^2\alpha_2^2$ & & 112211 & & 112211 & & & & & \\
$\alpha_1\alpha_2^3$ & & & & & & 111212 & & 111212 & \\
\end{tabular}
\end{ruledtabular}
\end{table}
\begin{table}[h!]
\vspace{-0.5cm}
\caption{\label{table C4} $a_{ijklmnpq}$ from Eq.~(\ref{Eq. S1}) for a 2D square crystal.}
\begin{ruledtabular}
\begin{tabular}{c|c c c c c c c c c}
\backslashbox{$\alpha_k\alpha_l\alpha_m\alpha_n\alpha_p\alpha_q$$~~~$}{$\beta_i\beta_j$}& & $\beta_1^2$ & & $\beta_2^2$ & & $\beta_1\beta_2$ & & $\beta_2\beta_1$ & \\
 \hline
$\alpha_1^6$ & & 11111111 & & 11111122 & &  & & & \\
$\alpha_2^6$ & & 11111122 & & 11111111 & &  & & & \\
$\alpha_1^4\alpha_2^2$ & & 11112211 & & 11222211 & & & & & \\
$\alpha_1^2\alpha_2^4$ & & 11222211 & & 11112211 & & & & & \\
$\alpha_1^5\alpha_2$ & & & & & & 11111212 & & 11111212 & \\
$\alpha_1^3\alpha_2^3$ & & & & & & 11122212 & & 11122212 & \\
$\alpha_1\alpha_2^5$ & & & & & & 11111212 & & 11111212 & \\
\end{tabular}
\end{ruledtabular}
\end{table}

The longitudinal resistivity can thus be written as
\begin{align}\label{Eq. S3}
\rho_L (\overrightarrow \alpha  ,\overrightarrow \beta  ) & = {a_{11}} + {a_{1111}}(\alpha _1^2\beta _1^2 + \alpha _2^2\beta _2^2) + {a_{1122}}(\alpha _1^2\beta _2^2 + \alpha _2^2\beta _1^2) + 4{a_{1212}}{\alpha _1}{\alpha _2}{\beta _1}{\beta _2} \notag\\
& + {a_{111111}}(\alpha _1^4\beta _1^2 + \alpha _2^4\beta _2^2) + {a_{111122}}(\alpha _1^4\beta _2^2 + \alpha _2^4\beta _1^2) + 6{a_{112211}}\alpha _1^2\alpha _2^2 + 8{a_{111212}}{\alpha _1}{\alpha _2}{\beta _1}{\beta _2} \notag\\
& + {a_{11111111}}(\alpha _1^6\beta _1^2 + \alpha _2^6\beta _2^2) + {a_{11111122}}(\alpha _1^6\beta _2^2 + \alpha _2^6\beta _1^2) + 15{a_{11112211}}(\alpha _1^4\alpha _2^2\beta _1^2 + \alpha _1^2\alpha _2^4\beta _2^2) \notag\\
 & + 15{a_{11222211}}(\alpha _1^4\alpha _2^2\beta _2^2 + \alpha _1^2\alpha _2^4\beta _1^2) + 12{a_{11111212}}(\alpha _1^5{\alpha _2} + {\alpha _1}\alpha _2^5){\beta _1}{\beta _2} + 40{a_{11122212}}\alpha _1^3\alpha _2^3{\beta _1}{\beta _2},
\end{align}
or, in terms of $\theta$ and $\varphi$,
\begin{equation}\label{Eq. S4}
\rho_L(\theta,\varphi) = C^\prime_0 + {C^\prime_2}\cos (2\theta )\cos (2\varphi ) + {S^\prime_2}\sin(2\theta )\sin(2\varphi ) + {C^\prime_4}\cos (4\theta ) + {C^\prime_6}\cos (6\theta )\cos (2\varphi ) + {S^\prime_6}\sin(6\theta )\sin(2\varphi ),
\end{equation}
where
\begin{align*}
C_0^\prime &= {a_{11}} + \frac{1}{2}\left( {{a_{1111}} + {a_{1122}}} \right) + \frac{3}{8}\left( {{a_{111111}} + {a_{111122}} + 2{a_{112211}}} \right) + \frac{5}{{16}}\left( {{a_{11111111}} + {a_{11111122}} + 3{a_{11112211}} + 3{a_{11222211}}} \right),\\
{C^\prime_2} & = \frac{1}{2}\left( {{a_{1111}} - {a_{1122}}} \right) + \frac{1}{2}\left( {{a_{111111}} - {a_{111122}}} \right) + \frac{{15}}{{32}}\left( {{a_{11111111}} - {a_{11111122}} + {a_{11112211}} - {a_{11222211}}} \right),\\
{S^\prime_2} & = {a_{1212}} + 2{a_{111212}} + \frac{{15}}{8}\left( {{a_{11111212}} + {a_{11122212}}} \right),\\
{C^\prime_4} & = \frac{1}{8}\left( {{a_{111111}} + {a_{111122}} - 6{a_{112211}}} \right) + \frac{3}{{16}}\left( {{a_{11111111}} + {a_{11111122}} - 5{a_{11112211}} - 5{a_{11222211}}} \right),\\
{C^\prime_6} & = \frac{1}{{32}}\left( {{a_{11111111}} - {a_{11111122}} - 15{a_{11112211}} + 15{a_{11222211}}} \right),\\
{S^\prime_6} &= \frac{1}{8}\left( {3{a_{11111212}} - 5{a_{11122212}}} \right).\\
\end{align*}
Thus, the AMR can be expressed as
\begin{equation}\label{Eq. S5}
\mathrm{AMR}_\mathrm{sq} = {{C}_2}\cos (2\theta )\cos (2\varphi ) + {{S}_2}\sin(2\theta )\sin(2\varphi )
 + {{C}_4}\cos (4\theta ) + {{C}_6}\cos (6\theta )\cos (2\varphi ) + {{S}_6}\sin(6\theta )\sin(2\varphi ),
\end{equation}
where ${C}_2$, ${S}_2$, ${C}_4$, ${C}_6$, and ${S}_6$ are, respectively, $C^\prime_2$, $S^\prime_2$, $C^\prime_4$, $C^\prime_6$, and $S^\prime_6$ divided by mean resistivity $C_0^\prime$. The corresponding expression for the PHE is
\begin{equation}
\mathrm{PHE}_\mathrm{sq} = - {{C}_2}\cos (2\theta )\sin(2\varphi ) + {{S}_2}\sin(2\theta )\cos(2\varphi ) - {{C}_6}\cos (6\theta )\sin (2\varphi ) + {{S}_6}\sin(6\theta )\cos(2\varphi ).
\end{equation}

\subsection{2D hexagonal crystal}
For a 2D hexagonal crystal (space group \textbf{p6m} / point group $D_6$), the non-vanishing coefficients in Eq.~(\ref{Eq. S1}) and the relations between them 
are given in Tables \ref{table C1}, \ref{table C2}, \ref{table C5}, and \ref{table C6}.
As in the square case, only coefficients with an even number of indices equal to 1 and even number equal to 2 do not vanish due to \emph{mirror} symmetry. 
In addition, 60$^\circ$ \textit{rotation} symmetry implies that \cite{Birss}:
\begin{align}
{a_{ijkl...}} = {\sigma _{ip}}{\sigma _{jq}}{\sigma _{kr}}{\sigma _{ls}}...{a_{pqrs...}},
\end{align}
where
\begin{align}
{\sigma} = \left( {\begin{array}{*{20}{c}}
{ \frac{1}{2}}&{-\frac{{\sqrt 3 }}{2}}\\
{ \frac{{\sqrt 3 }}{2}}&{ \frac{1}{2}}
\end{array}} \right).
\end{align}
which leads to the following relations:
\begin{align}\label{Eq.a1}
{a_{1111}} &= {a_{1122}} + 2{a_{1212}},
\end{align}
as well as
\begin{align}\label{Eq.a2}
{a_{111122}} &= 2{a_{111111}} + 6{a_{112211}} - 3{a_{222222}}, \nonumber\\
{a_{112222}} &= {a_{111111}} + {a_{112211}} - {a_{222222}}, \nonumber\\
{a_{222211}} &= 3{a_{111111}} + 6{a_{112211}} - 4{a_{222222}}, \nonumber\\
{a_{111212}} &=  - {a_{111111}} - \frac{3}{2} {a_{112211}} + \frac{3}{2} {a_{222222}}, \nonumber\\
{a_{122212}} &=  - \frac{3}{2}{a_{112211}} + \frac{1}{2}{a_{222222}},
\end{align}
\begin{table}[ht!]
\vspace{-0.5cm}
\caption{\label{table C5} $a_{ijklmn}$ from Eq.~(\ref{Eq. S1}) for a 2D hexagonal crystal.} 
\begin{ruledtabular}
\begin{tabular}{c|c c c c c c c c c}
\backslashbox{$\alpha_k\alpha_l\alpha_m\alpha_n$}{$\beta_i\beta_j$} & & $\beta_1^2$ & & $\beta_2^2$ & & $\beta_1\beta_2$ & & $\beta_2\beta_1$ & \\ \hline
$\alpha_1^4$ & & 111111 & & 111122 & &  & & & \\
$\alpha_2^4$ & & 222211 & & 222222 & &  & & & \\
$\alpha_1^3\alpha_2$ & & & & & & 111212 & & 111212 & \\
$\alpha_1^2\alpha_2^2$ & & 112211 & & 112222 & & & & & \\
$\alpha_1\alpha_2^3$ & & & & & & 122212 & & 122212 & \\
\end{tabular}
\end{ruledtabular}
\end{table}
\begin{table}[ht!]
\vspace{-0.5cm}
\caption{\label{table C6} $a_{ijklmnpq}$ from Eq.~(\ref{Eq. S1}) for a 2D hexagonal crystal.} 
\begin{ruledtabular}
\begin{tabular}{c|c c c c c c c c c}
\backslashbox{$\alpha_k\alpha_l\alpha_m\alpha_n\alpha_p\alpha_q$$~~~$}{$\beta_i\beta_j$}& & $\beta_1^2$ & & $\beta_2^2$ & & $\beta_1\beta_2$ & & $\beta_2\beta_1$ & \\
 \hline
$\alpha_1^6$ & & 11111111 & & 11111122 & &  & & & \\
$\alpha_2^6$ & & 22222211 & & 22222222 & &  & & & \\
$\alpha_1^4\alpha_2^2$ & & 11112211 & & 11112222 & & & & & \\
$\alpha_1^2\alpha_2^4$ & & 11222211 & & 11222222 & & & & & \\
$\alpha_1^5\alpha_2$ & & & & & & 11111212 & & 11111212 & \\
$\alpha_1^3\alpha_2^3$ & & & & & & 11122212 & & 11122212 & \\
$\alpha_1\alpha_2^5$ & & & & & & 12222212 & & 12222212 & \\
\end{tabular}
\end{ruledtabular}
\end{table}

and
\begin{flalign}\label{Eq.a3}
15{a_{11222211}} &= 3{a_{11111111}} + 6{a_{11111122}} - 4{a_{22222211}} - 2{a_{22222222}}, \nonumber\\
15{a_{11222222}} &= 6{a_{11111111}} + 3{a_{11111122}} - 2{a_{22222211}} - 4{a_{22222222}}, \nonumber\\
15{a_{11112211}} &=  - 4{a_{11111111}} - 2{a_{11111122}} + 3{a_{22222211}} + 6{a_{22222222}}, \nonumber\\
15{a_{11112222}} &=  - 2{a_{11111111}} - 4{a_{11111122}} + 6{a_{22222211}} + 3{a_{22222222}}, \nonumber\\
12{a_{11111212}} &=  - {a_{11111111}} + {a_{11111122}} - 3{a_{22222211}} + 3{a_{22222222}}, \nonumber\\
20{a_{11122212}} &= {a_{11111111}} - {a_{11111122}} - {a_{22222211}} + {a_{22222222}}, \nonumber\\
12{a_{12222212}} &= 3{a_{11111111}} - 3{a_{11111122}} + {a_{22222211}} - {a_{22222222}}.
\end{flalign}


As a result of all this, the longitudinal resistivity can be written as
\begin{align}\label{Eq. S6}
\rho_L (\overrightarrow \alpha  ,\overrightarrow \beta  ) &= {a_{11}} + {a_{1111}}(\alpha _1^2\beta _1^2 + \alpha _2^2\beta _2^2) + {a_{1122}}(\alpha _1^2\beta _2^2 + \alpha _2^2\beta _1^2) + 4{a_{1212}}{\alpha _1}{\alpha _2}{\beta _1}{\beta _2} + {a_{111111}}\alpha _1^4\beta _1^2 + {a_{111122}}\alpha _1^4\beta _2^2 \notag\\
& + {a_{222211}}\alpha _2^4\beta _1^2 + {a_{222222}}\alpha _2^4\beta _2^2 + 6{a_{112211}}\alpha _1^2\alpha _2^2\beta _1^2 + 6{a_{112222}}\alpha _1^2\alpha _2^2\beta _2^2 + 8{a_{111212}}\alpha _1^3{\alpha _2}{\beta _1}{\beta _2} + 8{a_{122212}}{\alpha _1}\alpha _2^3{\beta _1}{\beta _2} \notag\\
& + {a_{11111111}}\alpha _1^6\beta _1^2 + {a_{11111122}}\alpha _1^6\beta _2^2 + {a_{22222211}}\alpha _2^6\beta _1^2 + {a_{22222222}}\alpha _2^6\beta _2^2 + 15{a_{11112211}}\alpha _1^4\alpha _2^2\beta _1^2 + 15{a_{11112222}}\alpha _1^4\alpha _2^2\beta _2^2 \notag\\
& + 15{a_{11222211}}\alpha _1^2\alpha _2^4\beta _1^2 + 15{a_{11222222}}\alpha _1^2\alpha _2^4\beta _2^2 + 12{a_{11111212}}\alpha _1^5{\alpha _2}{\beta _1}{\beta _2} + 40{a_{11122212}}\alpha _1^3\alpha _2^3{\beta _1}{\beta _2} + 12{a_{12222212}}{\alpha _1}\alpha _2^5{\beta _1}{\beta _2},
\end{align}
or in terms of $\theta$ and $\varphi$, 
\begin{equation}\label{Eq. S7}
\rho_L(\theta,\varphi)  = C^\prime_0 + {C^\prime_2}\left[ {\cos (2\theta )\cos (2\varphi ) + \sin(2\theta )\sin(2\varphi )} \right] + {C^\prime_4}\left[ {\cos (4\theta )\cos (2\varphi ) - \sin(4\theta )\sin(2\varphi )} \right] + {C^\prime_6}\cos (6\theta ),
\end{equation}
where
\begin{align*}
C^\prime_0 & = {a_{11}} + \frac{1}{2}\left( {{a_{1111}} + {a_{1122}}} \right) + \frac{1}{2}\left( {3{a_{111111}} + 6{a_{112211}} - 3{a_{222222}}} \right) + \frac{1}{4}\left( {{a_{11111111}} + {a_{11111122}} + {a_{22222211}} + {a_{22222222}}} \right),\\
{C^\prime_2} & = \frac{1}{4}\left( {{a_{1111}} - {a_{1122}} + 2{a_{1212}}} \right) + \frac{1}{2}\left( { - 2{a_{111111}} - 6{a_{112211}} + 4{a_{222222}}} \right) + \frac{1}{4}\left( {{a_{11111111}} - {a_{11111122}} - {a_{22222211}} + {a_{22222222}}} \right),\\
{C^\prime_4} & = \frac{1}{2}\left( {{a_{111111}} - {a_{222222}}} \right) + \frac{1}{4}\left( {{a_{11111111}} - {a_{11111122}} + {a_{22222211}} - {a_{22222222}}} \right),\\
{C^\prime_6} & = \frac{1}{4}\left( {{a_{11111111}} + {a_{11111122}} - {a_{22222211}} - {a_{22222222}}} \right).
\end{align*}
Thus, the AMR can be expressed as
\begin{equation}\label{Eq. S8}
\mathrm{AMR}_\mathrm{hex} = {{C}_2}\left[ {\cos (2\theta )\cos (2\varphi ) + \sin(2\theta )\sin(2\varphi )} \right]
 + {{C}_4}\left[ {\cos (4\theta )\cos (2\varphi ) - \sin(4\theta )\sin(2\varphi )} \right]
 + {{C}_6}\cos (6\theta )
\end{equation}
where ${C}_2$, ${C}_4$, and ${C}_6$ are, respectively, $C^\prime_2$, $C^\prime_4$, and $C^\prime_6$ divided by the mean resistivity $C_0^\prime$. Fig.~\ref{FigureS3} presents the $AMR_{IP}$ data for three different devices fitted to above expression (See Fig. 5 and the manuscript for more details). The corresponding expression for the PHE is
\begin{equation}
\label{Eq. S9}
\mathrm{PHE}_\mathrm{hex} = -{{C}_2}\left[ {\cos (2\theta )\sin (2\varphi ) - \sin(2\theta )\cos(2\varphi )} \right]
 - {{C}_4}\left[ {\cos (4\theta )\sin (2\varphi ) + \sin(4\theta )\cos(2\varphi )} \right].
\end{equation}
Interestingly, the six-fold term drops out of the PHE in a hexagonal crystal.

\begin{figure}
\begin{center}
\includegraphics[width=0.35\hsize]{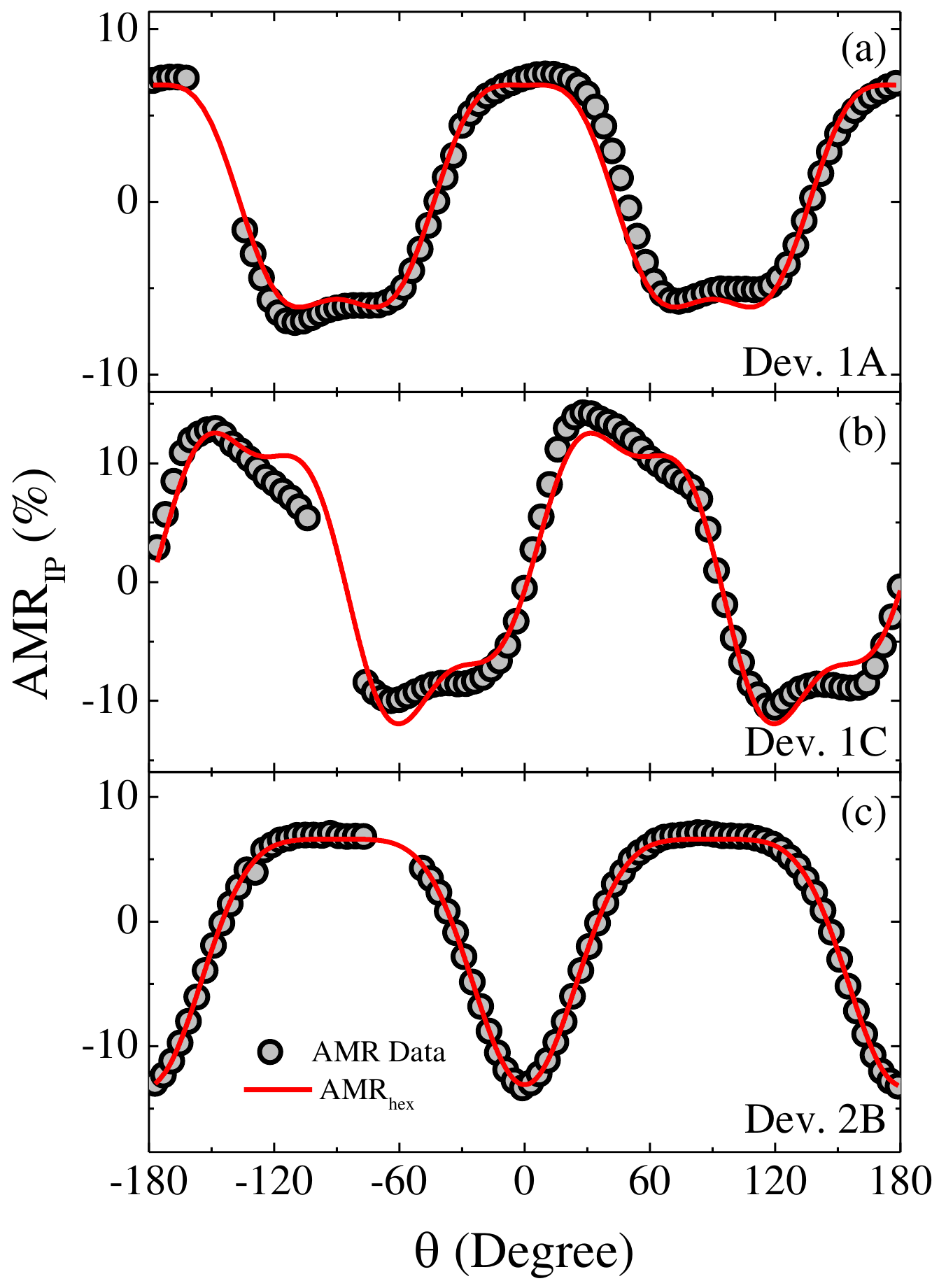}
\caption {AMR$_{IP}$ for three devices along with the fits using the AMR expression [Eq.~(\ref{Eq. S8}) or Eq. (1) in the main text] for 2D hexagonal crystal. The data presented here is same as shown in Fig. 5 of the manuscript. \label{FigureS3}}
\end{center}
\end{figure}

\begin{figure}[h]
\begin{center}
\includegraphics[width=0.4\hsize]{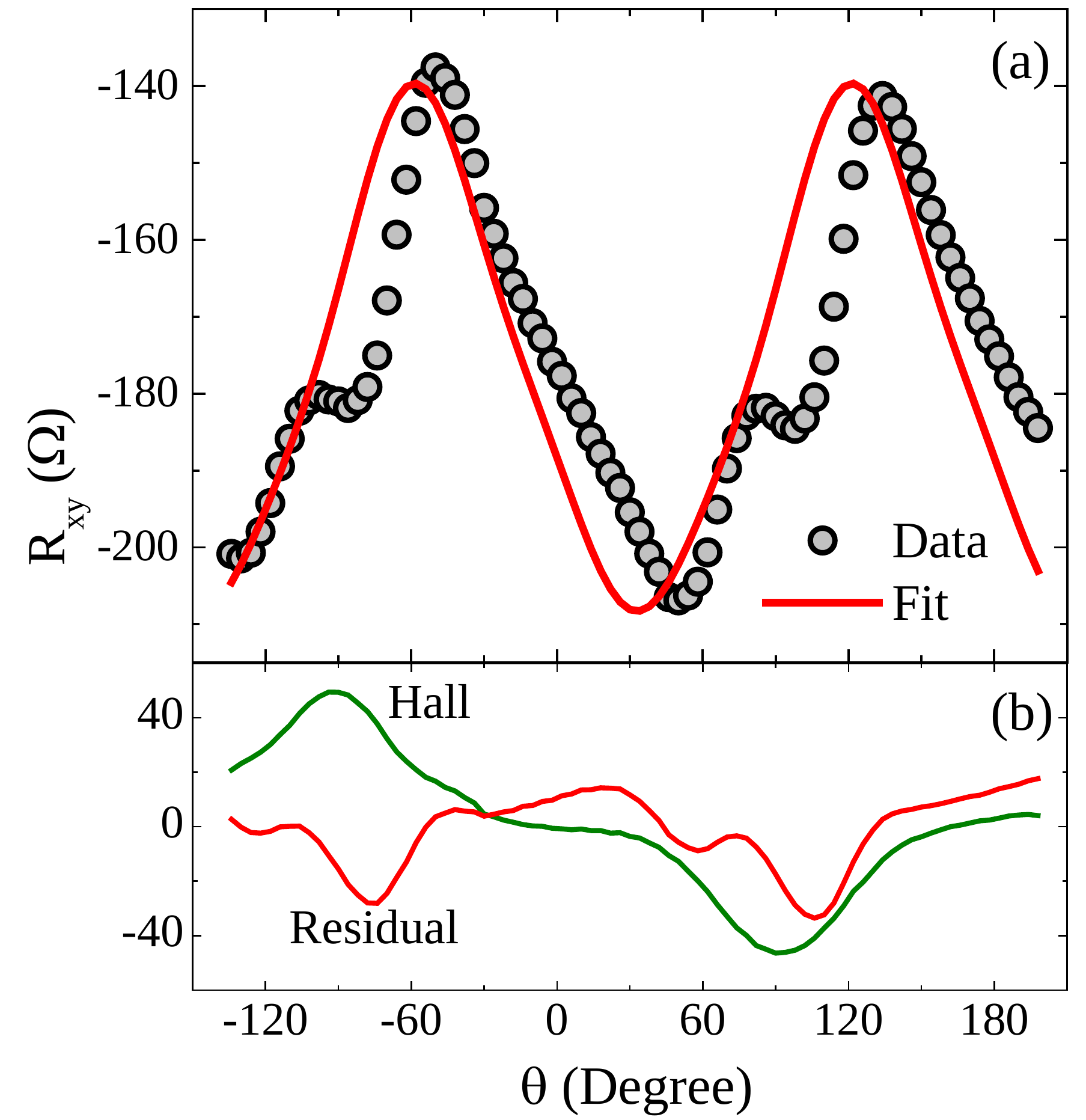}
\caption {(a) The transverse resistance $R_{xy}$ of device 1A measured at $H =$ 13.5~T, $V_g =$ 40~V and $T =$ 5~K. The solid line is a best fit to $[\mathrm{PHE_{hex}} \cos \epsilon + (1+\mathrm{AMR_{hex}}) \sin \epsilon]R_{L,\mathrm{mean}}$ with $\epsilon =$ -9.1$^{\circ}$ (see the text for more details). (b) The residual of the fit to the PHE data and the field-antisymmetric part of $R_{xy}$ (normal Hall contribution). \label{FigureS4}}
\end{center}
\end{figure}

\section{PHE measurement}

Fig.~\ref{FigureS4}(a) presents a typical PHE curve at $V_g =$ 40~V for device 1A after symmetrization of the data measured at +13.5~T and -13.5~T. The nonzero average value of the PHE indicates the presence of some longitudinal resistance due to contact misalignment. Therefore, the data can be assumed to be a linear combination of the PHE and AMR, $[\mathrm{PHE_{hex}} \cos \epsilon + (1+\mathrm{AMR_{hex}}) \sin \epsilon]R_{L,\mathrm{mean}}$, where AMR$_\mathrm{hex}$ and PHE$_\mathrm{hex}$ are given by Eqs.~(\ref{Eq. S8}) and~(\ref{Eq. S9}), respectively. $\epsilon$ represents the misalignment angle, whereas $R_{L,\mathrm{mean}}$ is the angle-averaged longitudinal resistance. Using the $R_{L,\mathrm{mean}}$ and $C_n$ obtained from the fit of the AMR data (cf.\ Fig.~4 of main text), we can get a reasonably good fit to the PHE data with a single fitting parameter $\epsilon$. We do see relatively large deviations near -80$^{\circ}$ and 100$^{\circ}$. However, these positions coincide with the region where the antisymmetric part with respect to the magnetic field (normal Hall contribution) of $R_{xy}$ is large [see Fig. \ref{FigureS4}(b)]. This Hall signal can arise from small perpendicular field due to a wobble of the probe. Hence, we believe that the observed deviation of the PHE is related to instrumental artifacts.
In principle, the perpendicular field could have also modified the AMR data. However, extracting the perpendicular magnetic field from the Hall data in Fig.~\ref{FigureS4}(b), we find the corresponding perpendicular MR values to be less than 0.2$\%$. Therefore, the AMR data is practically insensitive to the wobble, and for this reason we concentrated on it for the analysis presented in the main text.

\begin{figure}[h]
\begin{center}
\includegraphics[width=0.6\hsize]{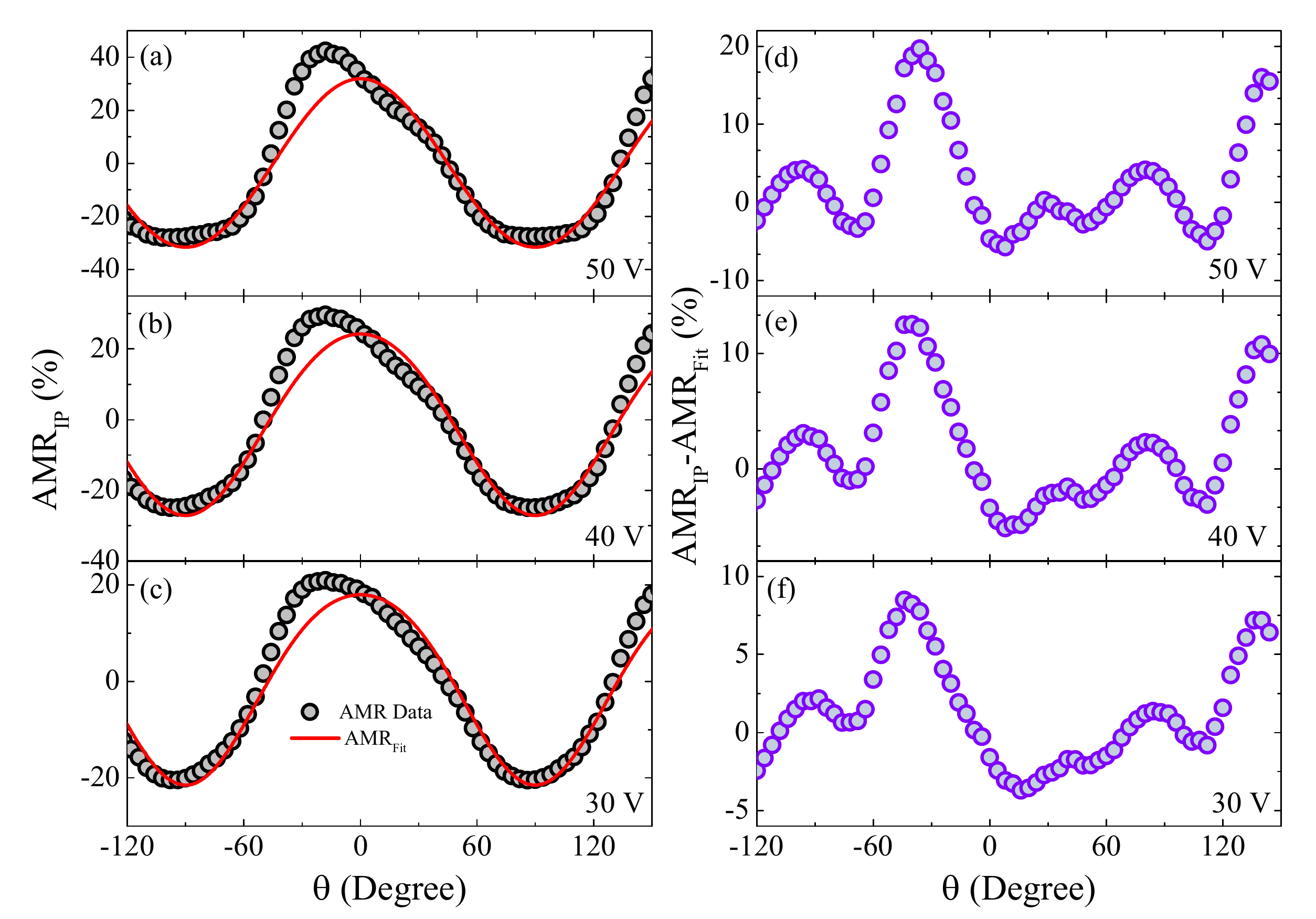}
\caption {AMR$_{IP}$ measured on 100~$\mu$m $\times$ 260~$\mu$m Hall bar device (sample I3) oriented along [1$\bar 1$0] at 5~K for $V_g =$ 50~V (a), 40~V (b), and 30~V (c). The corresponding residuals are shown in (d-f).\label{FigureS5}}
\end{center}
\end{figure}

\begin{figure}[h]
\begin{center}
\includegraphics[width=0.6\hsize]{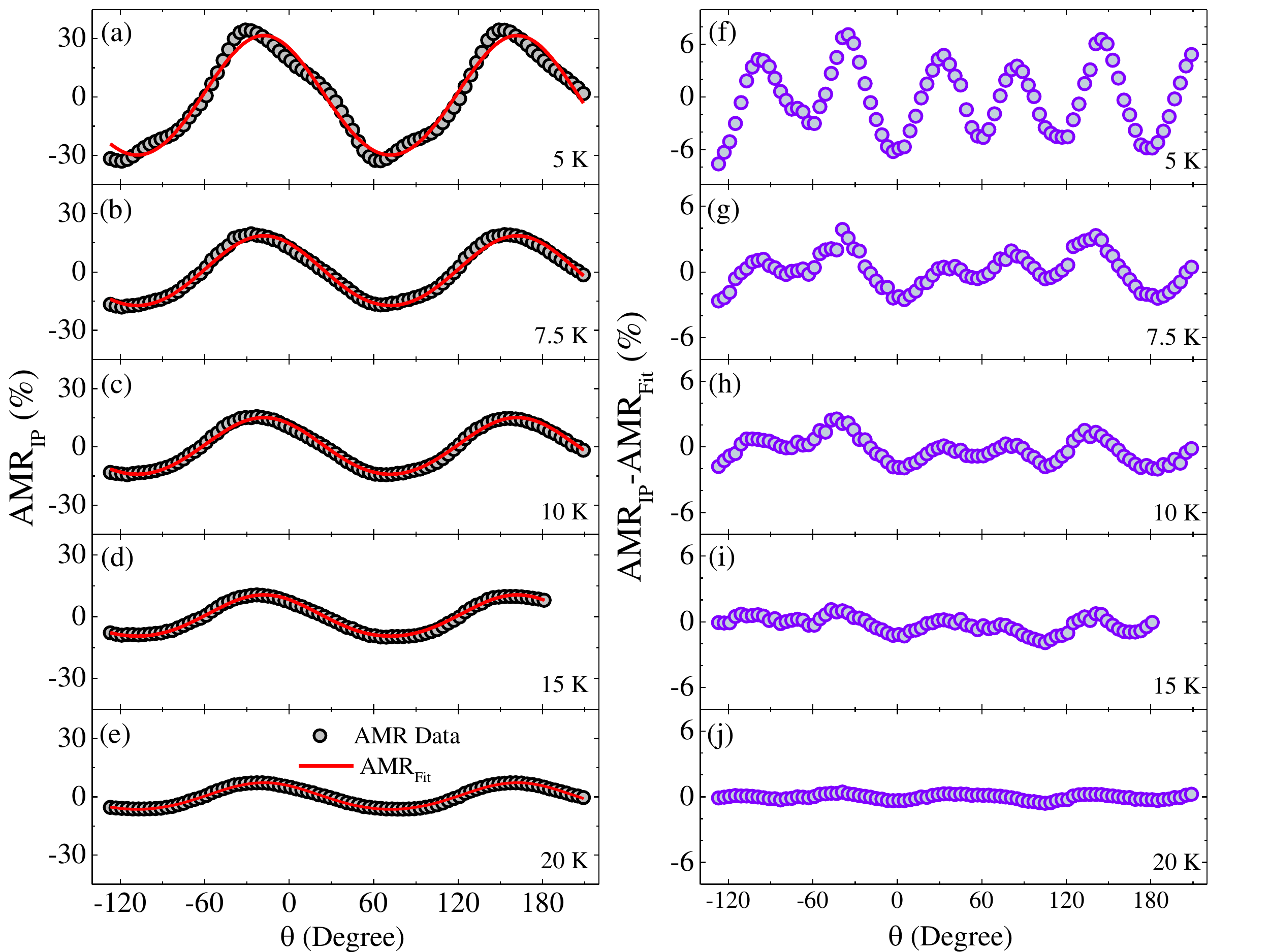}
\caption {The AMR$_{IP}$ measured on a smaller 10~$\mu$m $\times$ 28~$\mu$m Hall bar (sample I4) oriented along [1$\bar 1$0] at $V_g =$ 0~V for $T =$ 5~K (a), 7.5~K (b), 10~K (c), 15~K (d), and 20~K (e). The corresponding residual are presented in (f-j).\label{FigureS6}}
\vspace{-0.5cm}
\end{center}
\end{figure}


\section{Additional in-plane AMR measurements}

We present the in-plane AMR data of two more LAO/STO (111) devices. Sample I3 is similar to the sample mentioned in the main text, and was accordingly prepared using a three step deposition process (as described in Ref.~\cite{Maniv}), while a single step deposition process was employed to fabricate a smaller (10~$\mu$m $\times$ 28~$\mu$m) Hall bar devices on sample I4~\cite{Maniv2}.

Fig.~\ref{FigureS5} displays the AMR$_{IP}$ for a device on sample I3 which is oriented along [1$\bar 1$0] for positive gate voltages. We can see equispaced peaks separated by 60$^\circ$ in the residual $AMR_{IP}-AMR_{Fit}$ due to hexagonal symmetry, which is quite similar to the results shown in the main text.
The AMR$_{IP}$ data of the smaller device on sample I4 oriented along [1$\bar 1$0] for various temperatures are presented in Fig.~\ref{FigureS6}. It displays a similar behavior to the larger device, thus ruling out structural domains as the origin of the six-fold effect, as explained in the main text.
The temperature dependence of the AMR coefficients for the last device is displayed in Fig.~4(f) of the manuscript.

\section{Effect of thermal cycling}

The thermal history dependence of the AMR along [1$\bar 1$0] in sample I4 is presented in Fig.~\ref{FigureS7}. We observe six-fold AMR components for two different cool-downs. However the position of the uniaxial component is different for both cases. This further strengthens our proposed scenario that relates the uniaxial component of the AMR to the tetragonal domain structure, which should indeed reorganize upon heating above the cubic-to-tetragonal transition at 105 K and cooling back down.

\begin{figure}[h]
\begin{center}
\includegraphics[width=0.6\hsize]{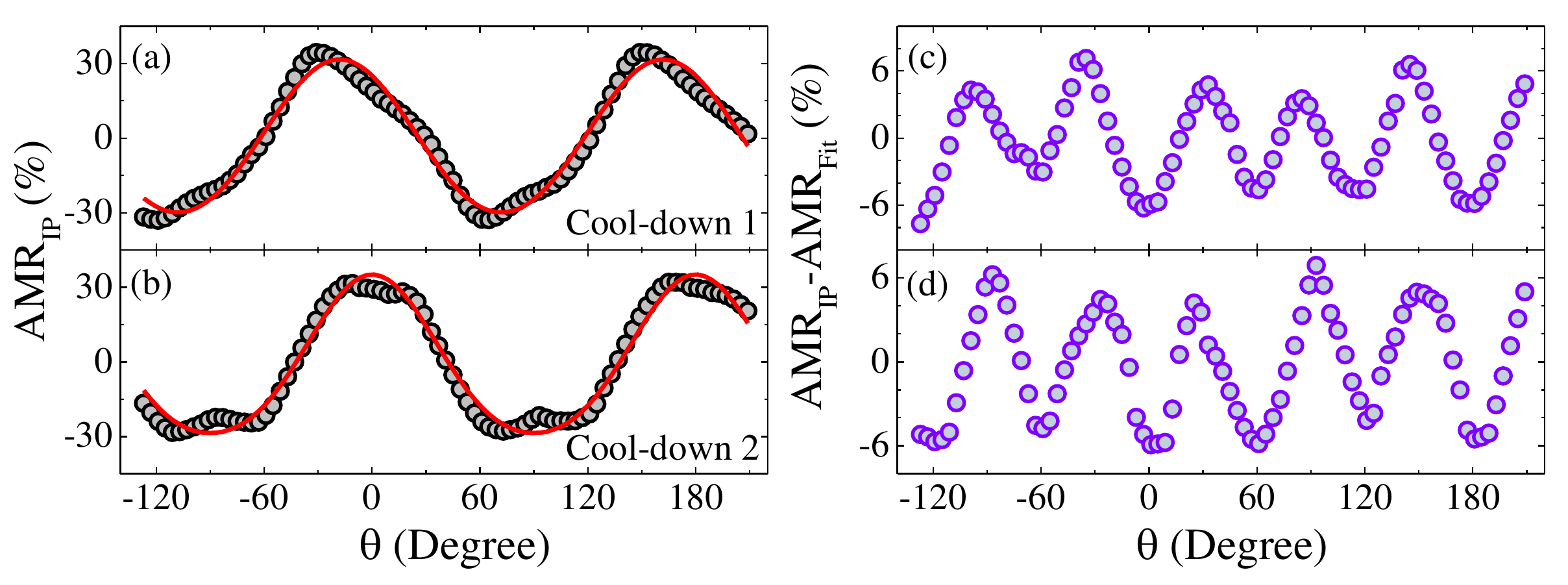}
\caption {The AMR$_{IP}$ measured on a device oriented along [1$\bar 1$0] for two different cool-downs from 300 K. The measurements are done at $V_g =$ 0~V and $T =$ 5~K. The residuals corresponding to (a-b) are presented in (c-d).\label{FigureS7}}
\vspace{-0.5cm}
\end{center}
\end{figure}

\vspace{-0.25cm}
\bibliographystyle{apsrev}

\end{widetext}

\end{document}